\documentclass{aastex631}

\usepackage{graphicx}

\shortauthors{Wright \etal}

\shorttitle{NEOWISE on 2001 CC$_{21}$}

\newcommand{\etal}         {{\it et al.}}
\newcommand{\vs}           {{\it vs.}}
\newcommand{\asec}      {\mbox{$^{\prime\prime}$}}

\newcommand{\um}           {\mbox{$\mu$m\ }}

\newcommand{\be}           {\begin{equation}}
\newcommand{\ee}           {\end{equation}}
\newcommand{\bea}          {\begin{eqnarray}}
\newcommand{\eea}          {\end{eqnarray}}

\begin{document}

\title{NEOWISE data and Thermophysical Modeling of 98943 Torifune (2001 CC$_{21}$)}

\correspondingauthor{Edward L. (Ned) Wright}
\email{wright@astro.ucla.edu, jmasiero@ipac.caltech.edu,, mainzer@epss.ucla.edu}
\author[0000-0001-5058-1593]{Edward L. (Ned) Wright}
\affiliation{UCLA Dept.\ of Physics \& Astronomy\\
P.O. Box 951547\\
Los Angeles, CA 90095-1547\\
USA}
\author{Joseph Masiero}
\affiliation{IPAC}
 \author{Amy Mainzer}
\affiliation{UCLA Dept. of Earth, Planetary \& Space Sciences}

\section{Abstract}

The Hayabusa2$\sharp$ flyby target 98943 Torifune (2001 CC$_{21}$) has an uncertain size
based on an uncertain albedo and uncertain absolute magnitude.
We have collected all the NEOWISE observations of 2001 CC$_{21}$
from Nov2021 through Feb 2024, a total of 132 frames, 
and analyzed this data to estimate an
infrared radiometric diameter.
We  analyze the multi-epoch 3.4 \um\ \& 4.6 \um\ NEOWISE data using
an ellipsoidal  rotating, cratered ThermoPhysical Model (TPM) to obtain estimates
for the diameter, rotation pole, shape, and thermal inertia.
2001 CC$_{21}$ is quite faint at 4.6 \um\ when $\Delta \sim 0.7$
AU, so the resulting diameter is substantially smaller than the
700 m derived from the $H$ magnitude and L spectral type.
Recent polarimetric data has also suggested a smaller diameter, but
not quite as small as the diameter derived from the thermal IR data.
A fit to an ellipsoidal TPM model gives a volume equivalent
sphere diameter of $337_{-27}^{+33}$
meters [posterior median and central 68\% confidence interval].  
Prograde rotation with an obliquity of $(24_{-9}^{+6})^\circ $ is preferred.
We also applied this TPM to the Spitzer data
presented by Fornasier \etal\ (2024) and obtain a diameter of 476$\pm  9\%$ meters
which is consistent with the NEATM modeling presented by Fornasier \etal\ but
with more realistic errorbars.
Finally, fitting the NEOWISE and Spitzer data together 
requires unexpectedly large thermal inertias and gives a bimodal 
posterior diameter distribution.

\section{Introduction}

The Apollo asteroid 98943 Torifune
(hereafter called 2001 CC$_{21}$, its provisional designation) has an orbital period of 1.05
years and thus 20 orbits of the asteroid corresponds to very nearly
21 Earth years.  
This gives a long synodic period, so this object
has 3-4 oppositions in consecutive years followed by 17-18 years
without oppositions.
After its discovery in 2001, 2001 CC$_{21}$ was observed quite often
until December 2004.  
It was then observed again starting in September 2019 until the present.
Two precovery observations from 10 Nov 1982 have contributed 
to a well-determined orbit.

Popescu \etal\ (2025) have analyzed optical lightcurve data from 2001-2002
and 2022-2023 to determine an accurate rotation period of 5.02 hours
and approximate rotation pole and convex shape model.  The
favored rotation pole gives prograde rotation.   
Combining lightcurves taken at different phase angles Popescu \etal\
also derived an optical absolute magnitude of $H = 18.78 \pm 0.14$ mag.

Spectrophotometric observations led to an SMASS classification of
L type (Binzel \etal\ 2004).  Hirabayashi \etal\ (2021), in discussing
the chosen Hayabusa2$\sharp$ mission, cite the L spectral type from 
Binzel \etal\ (2004) and give a diameter of 700\,$\mbox{m}$.
Mainzer \etal (2011) found a median albedo of 0.178 for L-type
asteroids, and with the absolute magnitude $H=18.6$ in Binzel \etal\ (2004)
the standard formula
\be
D = \frac{(1329\,\mbox{km})}{10^{H/5}\sqrt{p_V}}
\label{eq:DfromHp}
\ee 
(Bowell \etal\ 1989)
gives $D = 600\,\mbox{m}$, which suggests that Hirabayashi \etal\ (2021)
assumed an albedo of 0.13 to get a diameter of 700\,$\mbox{m}$.

Geem \etal\ (2023) performed polarimetric observations of 2001 CC$_{21}$
which can be used to estimate the albedo.  They found a higher albedo of
$p_V = 0.23 \pm 0.04$ and a diameter of 488 meters using an
absolute magnitude of $H = 18.77$.
\footnote{Note the arxiv preprint (v1) has a different albedo in the abstract and
the old H=18.6.}

Another diameter estimate  is given by Arimatsu \etal\ (20240
based on an occultation observations.  
Unfortunately only one chord of 400 m length was
measured, but this paper used a diffraction analysis to derive an elliptical
shape model for the asteroid projected on the sky with a semi-major axis
$a =  420^{+80}_{-60}$ m and an axis ratio $b/a = 0.37 \pm 0.09$.
This result is less precise than the usual multi-chord occultation 
observation.  Popescu \etal\ (2025) used this result to normalize their
shape model and give an equivalent ellipsoid with semi-axes of
$a =  420^{+80}_{-60}$, $b =  160^{+50}_{-40}$, and $c =  170 \pm 30$
meters.
We note the oddity that $c > b$ but within the errors this shape is
consistent with a prolate shape with minimal triaxiality.
Combining these axis lengths lets Popescu \etal\ (2025)  dervie
an equivalent spherical diameter of 
$2(abc)^{1/3} = 440 \pm 60$ meters.

Radiometric diameter estimates using thermal infrared data and either
thermal models or thermophysical models can be used to improve the
accuracy of the size and albedo.  Fornasier \etal\ (2024) analyze Spitzer IRS data 
taken in 2005 using the Near Earth Asteroid Thermal Model (Harris  1998),
and get a diameter of  $465 \pm 15$ meters.  Fornasier \etal\  also derived
a new, even fainter $H=18.94 \pm 0.05$ and an albedo of $0.216 \pm 0.016$.
Using the Geem \etal\ (2023) albedo with the Fornasier \etal\ $H$ gives
a diameter of $451.6^{+9.8\%}_{-8.3\%}$ meters

In this paper we use the NEOWISE (Near Earth Object WISE) 
data as input to a thermophysical model.
The WISE (Wide-field Infrared Survey Explorer) 
spacecraft (Wright \etal\ 2010) was launched in December 2009 and surveyed
the whole sky in 4 infrared bands 
(3.4, 4.6, 11, \& 22 \um) between 7 January 2010 and 6 August  2010, then continued 
without the 22 \um\ band through September  2010, and with only the 3.4 and 4.6 \um\
bands through Jan 2011.
The WISE spacecraft was reactivated in December  2013 as NEOWISE
(Mainzer \etal\ 2014b) and continued to observe through July 2024.
The data on
2001 CC$_{21}$ starts when it first went through the NEOWISE 
scan path in November 2021 and ends in February 2024.

Wright, Masiero \& Mainzer (2023) presented a preliminary version of this analysis, without the
last two epochs of NEOWISE data, which gave a diameter $329^{+78}_{-41}$ meters

\section{Summary of the NEOWISE data}

\begin{figure}[p]
\plotone{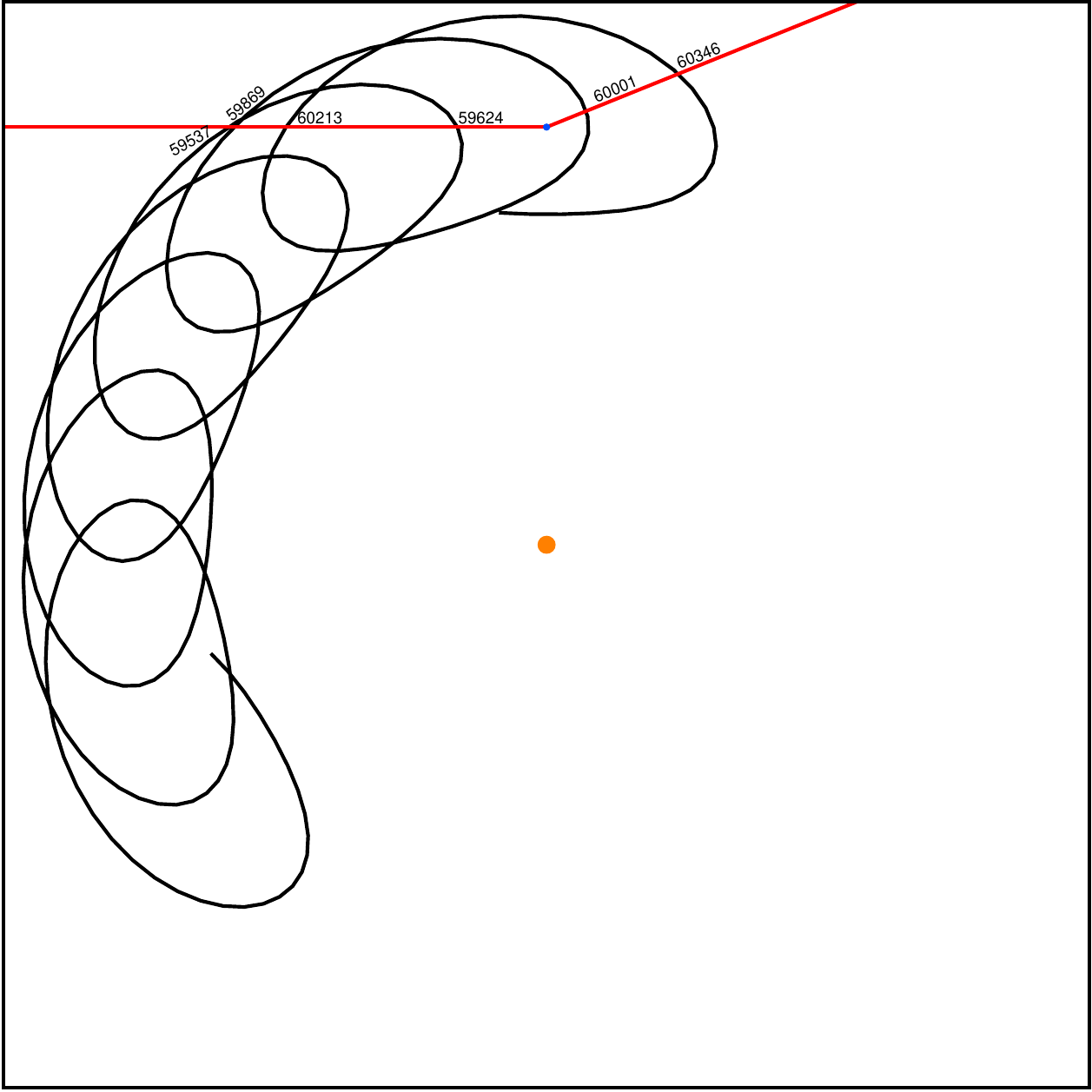}
\caption{Orbit of 2001 CC$_{21}$ relative to the Earth (blue dot)
in a frame rotating to keep the Sun on the $-y$ axis. 
The orange circle at $y = -1$ AU indicates the Sun's direction.
The plot covers the period from MJD 57754 (1 Jan 2017) 
in the lower left to 60484 (23 Jun 2024).
Lines indicating the current scan paths of the NEOWISE
mission are shown.  
Tiny labels give the MJD for each observing epoch..\label{fig:orbit}}
\end{figure}

The orbit of 2001 CC$_{21}$ is close to the orbit of the 
Earth, but with a slightly different period and higher eccentricity.
As a result its position relative to the Earth executes an epicyclic
motion around a slowly drifting center, as shown in 
Figure \ref{fig:orbit}.  Positions from 1 Jan 2017 to 23 Jun 2024
are shown.  NEOWISE stopped observing on 31 Jul 2024 and
burnt up in the atmosphere on 2 Nov 2024.  NEOWISE was
scanning at ecliptic longitudes of $-90^\circ$ and $+112^\circ$
from the Sun during the period of these observations.
The latter angle is chosen to reduce thermal
loading from the Earth.   These scans paths are shown in the 
Figure.  One can see 6 intersection points between the orbit of
2001 CC$_{21}$ and the NEOWISE scans.  
Each crossing leads to 
a set of frames that cover the predicted positions of 2001 CC$_{21}$,
and IRSA has a convenient option in the WISE image server to
download all of these frames.  The spacing of frames covering
2001 CC$_{21}$ is typically  once every orbit or two orbits when it is
close to the scan path, separated by gaps of several months
when it is away from the scan paths.
Sorting the frames by date and separating into sets with gaps
of at least 3 days gives epochs
centered on MJDs 59530.660, 59624.329, 59866.428, 59872.268,
60000.841, 60213.380, \& 60346.162, with 
$\Delta t$s of 0.7693, 0.2285, 0.7711, 0.7611,
0.1326, 1.8776, \&   0.9669 days.  
A maneuver around MJD 59869 to avoid the $3^{rd}$ quarter
Moon on 17 October 2022 split one of the 6 crossings into two epochs.

All of the frames within each epoch are then stacked to make a coadded
image of the moving object.  This stacking is done using medians
rather than means in order to allow for outliers, which can be
caused by fixed stars, artifacts, or cosmic rays.   Using the median
is equivalent to fitting data to a model using a least sum of
absolute value of deviations technique, which minimizes the L1 norm
of the deviations, as opposed to a least squares fit which minimizes
the L2 norm of the deviations.

This analysis that uses all the frames is more inclusive than an
analysis that uses only the detections in the single frame 
detection database.  For faint objects, the single frame detections
are often enhanced by noise peaks or celestially fixed WISE sources that 
just happen to be close to the predicted position of the asteroid, 
so frames with nearby WISE catalog sources were deleted before the 
stacking process.
Due to the lightcurve amplitude,
the difference between the flux in the stacked image and the mean
of only the detections can be substantial for a faint object like
2001 CC$_{21}$ at $\Delta = 0.7$ AU.

During the process of coadding the frames, a two parameter L1 norm
fit is done to a circular patch of pixels with radius of 18\asec\
in each frame.  
The pixel values are rescaled to make the magnitude corresponding
to one DN match the standard 20.5 in W1 and 19.5 in W2.
The circle is centered on the predicted position
of the moving object, and the two parameters are a constant background
and the scaling of a PSF model centered on the predicted source
position.   The scaling of the PSF model gives a source flux for
each frame, based on forced photometry at the predicted position.
Table \ref{tab:singles} gives the MJD, fitted flux, and the Median Absolute
Deviation of the fit for each frame.
These single frame fluxes and the MJD of the frame can be used for
finding rotational periods and amplitudes.  Since 2001 CC$_{21}$ has a
well determined rotational period, no period search was done.  The
amplitudes are based on fitting a sine wave with two maxima and two
minima per period to the data.    The data are fit to $F(t) =
A+B\sin(4\pi t/P)+ C\cos(4\pi t/P)$ and then the reported amplitude
is $(5/\ln10)\sqrt{B^2+C^2}/A$, where the scaling makes the amplitude
comparable to the usual max-min magnitudes.  No attempt is made to
maintain phase between the epochs.  Amplitudes are noise-biased,
and using a fit to many frames reduces the bias.  NEOWISE
data are often of lower SNR and sparser time sampling than optical
light-curve photometry, so reducing the noise bias is an important
consideration.

\begin{table}[tbp]
\begin{tabular}{rrrrr}
\multicolumn{1}{c}{MJD} & \multicolumn{1}{c}{Flux} &
\multicolumn{1}{c}{MAD} & \multicolumn{1}{c}{Frame ID} &
\multicolumn{1}{c}{Notes} \\
 \tableline
59529.44243  &   32.6  &    2.9  & 33634r135  & \\
59529.44256  &   0.9   &   2.6  & 33634r136 & \\
59529.57321  &   31.1  &    2.9  & 33638r135 & \\
59529.57333  &    3.0   &   3.0  & 33638r136 & \\
59529.70398  &   59.7   &   3.1  & 33642r136 & \\
59529.83463  &   4.4   &    2.7  & 33646r136 & \\
59529.96528  &   14.8  &    3.3  & 33650r135 & \\
59530.09606  &    3.7   &   3.1  & 33654r136 & \\
59530.22670  &   47.5  &    3.0  & 33658r136 & \\
59530.35735  &   46.4  &    2.9  & 33662r136 & \\
\multicolumn{4}{c}{\ldots} \\
\tableline
\end{tabular}
\tablecomments{Sample of single frame fluxes of 2001 CC21 in the W2 band
from a 2 parameter fit that minimizes the L1 norm.   
Flux and the Median Absolute Deviation (MAD) are in
rescaled DN such that 1 DN corresponds to W2 = 19.5.  
The Notes column contains
a D if the frame is deleted from the multiframe fits.  
Blank lines in table separate the epochs in Table \ref{tab:NEOWISE}.
This table is available in full as machine readable datai and also can be found after the 
references in this preprint.}

\caption{Single frame fluxes of 2001 CC21 in the W2 band.
\label{tab:singles}}
\end{table}

Finally all the pixels within 18\asec\ of the predicted source
position on any of the $N_F$ frames are fit to a model with $N_F+1$
parameters: $N_F$ parameters for frame-dependent backgrounds and a
single parameter for the source flux.  Again, this fit was done
using the least sum of absolute deviations, or the L1 norm.  These
multiframe based fluxes have been turned into magnitudes and are
reported in Table \ref{tab:NEOWISE}.  

Since the multiframe fluxes $\langle F \rangle$ can be negative, 
especially for W1 which has much lower SNR,
the magnitudes are given by $m = zp-2.5 \log_{10} (\vert\langle F \rangle\vert)$.
The flux uncertainties were evaluated
using jackknife resampling on a frame-by-frame basis, doing $N_f$
fits that each left out one frame.  
The error on the magnitude is computed using $\sigma(m) = 
2.5 \log_{10}(1+\sigma(F)/\vert\langle F \rangle\vert)$.  
Finally, if the flux is negative,
the $\sigma$ is reported as negative, but none of the fluxes for
2001 CC$_{21}$ were negative.  This approach allows the entry
of low SNR data while continuing to use magnitudes.
The mean MJD for the data is also reported.

The uncertainty in the mean flux includes the noise introduced by
the random sampling of the lightcurve phase.  This is typically
a fairly small uncertainty, since even a lightcurve with a peak to valley
ratio of 1:0.6, an 0.55 magnitude amplitude,  has a standard
deviation of 0.14 and a mean flux of 0.8.  With a typical 12 frames
per apparition, the mean NEOWISE flux uncertainty is only 5\%.
The error on the amplitude is taken to be $\sqrt{8}$
times the magnitude error, which is appropriate for
random sampling of a sinusoidal lightcurve.

As a final defense against outliers, the MCMC code calculates
a robust $\chi^2$ which is $x^2$ for $|x| < 2$ but switches to
$4+4(|x|-2)$  where $x = (\mbox{obs}-\mbox{calc})/\sigma$.
This effectively switches from fitting an $L_2$ norm to an 
$L_1$ norm for large deviations, and downweights the large
discrepancies.

The $H$ magnitude and rotation period are also input data, so there
are 19 data points for the ellipsoidal
model, although the SNR for the W1 data is often too low to constrain
the model.  The error on the $H$ magnitude is taken to be 0.15 mag
which is better than the typical uncertainty of 0.3 mag in Veres \etal\ (2015).  The slope 
parameter $G$ is assumed to be 0.15 (Bowell \etal\ 2015), which then
gives a phase integral of 0.4 for correcting the geometric albedo $p$
to the Bond albedo $A$ needed for thermal modeling.  Since the
factor that enters the thermal modeling is $(1-A)$, and $A$ is small
($A = 0.16$) even for the high albedos seen here, assumptions about
the $H$ magnitude have little effect on the diameter.
The error on the period is set to zero since the optical light curve data 
quality is sufficient to get the correct cycle count over 20+ years.

\begin{table}[p]
\begin{tabular}{rrrrrrrrr}
\multicolumn{1}{c}{MJD} & \multicolumn{1}{c}{$\alpha$} &
\multicolumn{1}{c}{$\delta$} & $\Delta$ [AU] & 
\multicolumn{1}{c}{W1} &  \multicolumn{1}{c}{W2}  &
 \multicolumn{1}{c}{N$_F$}  & Amp & $\lambda-\lambda_\odot$ \\
 \tableline
59530.660 & 143.294 & 16.309 & 0.7693 & $18.565 \pm 0.511$ & $16.221 \pm 0.301$ & 30 
& \nodata & -90 \\
59624.329 & 238.490 &  1.784 & 0.2285 & $15.848 \pm 0.063$ & $13.253 \pm 0.027$ & 17 
& 0.429 & -90 \\
59866.428 & 113.362 & 19.404 & 0.7711 & $20.504 \pm 2.140$ & $ 16.839 \pm 0.481$ & 23 
& \nodata & -90 \\
59872.268 & 116.753 & 19.254 & 0.7611 & $18.162 \pm 0.672$ & $15.926 \pm 0.354$ & 21 
& \nodata & -90 \\
60000.841 &  87.488 & 62.827 & 0.1326 & $14.539 \pm 0.167$ & $12.339 \pm 0.060$ & 7 
& 0.249 & 112 \\
60213.380 & 93.358 & 17.526 & 0.6271 & $18.697 \pm 0.881$ & $15.838  \pm 0.373$ & 21 &  \nodata & -90 \\
60346.162 & 64.733 & 33.090 & 0.3468 & $16.390 \pm 0.196$ & $14.607  \pm 0.126$ &13 & 0.836 & 112 \\
\end{tabular}
\caption{Summary of the NEOWISE data on 2001 CC$_{21}$.
MJD is the mean Julian Day - 2400000.5, 
$\alpha$ and $\delta$ are the mean right ascension and declination during the epoch in degrees,
W1 and W2 are the NEOWISE magnitudes found using a fit to all the pixels near the source
on all the frames in the epoch, $N_F$ is the number of frames in the epoch,
Amp is the amplitude as defined in the text, and $\lambda-\lambda_\odot$ is the ecliptic longitude
of the object relative to the Sun.}
\label{tab:NEOWISE}
\end{table}

\section{Ellipsoidal Thermophysical Model Fit}



\begin{table}[p]
\begin{tabular}{llrlr}
Parameter & Prior & Posterior \\
\tableline
Rotation Pole & uniform in $4\pi$ & obliquity $(24^{+7}_{-8})^\circ$\\
\vspace{1mm}
Diameter & log uniform in $[1..10^6]$ m  & $337^{+9.4\%}_{-8.4\%}$ m \\
\vspace{1mm}
$p_V$ & see Wright (2016) & $0.405 \pm 17.2\%$  \\
\vspace{1mm}
Period & 5.02 h & 5.02 h \\
\vspace{1mm}
Thermal inertia $\Gamma$ &
$\log(\Gamma_{\mbox{MKS}}) = 2.509-0.352\log(D[\mbox{km}]) \pm 0.2$ & $448^{+44.5\%}_{-52.5\%}$ \\
\vspace{1mm}
crater fraction & uniform in [0..1] &  $0.514^{+0.307}_{-0.323}$ \\
\vspace{1mm}
IR:optical albedo ratio &  $p_{IR}/p_V = 1.755\pm34\%$   & $1.484^{+16\%}_{-17\%}$ \\
b/a & $(b/a)^3$ uniform in [0..1] & $0.804\pm 0.06$ \\
c/b & $(c/b)^4$ uniform in [0..1] & $0.796^{+0.129}_{-0.162}$ \\
\end{tabular}
\caption{Ellipsoidal Thermophysical Model Parameters}
\label{tab:ellipsoid}
\end{table}


The input data for the model fits are the  H magnitude from Fornasier \etal, 
the period from Popescu \etal\ (2025), and the infrared magnitudes and amplitudes
in Table \ref{tab:NEOWISE} with one modification: the error on the
W1 magnitude on MJD 59624.329 was increased to 0.2 to avoid
dominating the other W1 epochs.

The ellipsoidal model used here has 10 parameters: the
diameter, the albedo, the thermal inertia, the rotation period,
the ratio of the infrared albedo to the optical albedo, the crater
fraction that determines the surface roughness, the
right ascension and declination of the rotation pole, and the
axis ratios $b/a$ and $c/b$.  Since we want to estimate diameters
even when the data are very sparse, all of the parameters
are constrained by priors given in the second column of
Table \ref{tab:ellipsoid}.   Penalty functions that enforce these
priors are included in the posterior probability density.

The TPM model depends on the thermal inertia
$\Gamma$ which combines the thermal conductivity $\kappa$,
the density $\rho$, and the heat capacity $C$.
The model uses a prior on
$\Gamma = \sqrt{\kappa\rho C}$ based on Hung \etal (2022), 
with 
\be
\log(\Gamma_{\mbox{MKS}}) = 2.509-0.352\log(D[\mbox{km}]) \pm 0.2
\ee
where the $\pm 0.2$  is the intrinsic scatter needed to 
make $\chi^2 \sim 1$ per degree of freedom in the Hung \etal\ fit.  
The MKS units for $\Gamma$ are W$\sqrt{\mbox{sec}}$/m$^2$/K.
Hung \etal\ (2022) assumed that $\Gamma$ scaled like $T^{1.5}$
and report values scaled to 1 AU from the Sun.
The posterior distribution after fitting the NEOWISE data is nearly identical to the prior, 
which gives $\Gamma = 473 \pm 46\%$ at the posterior median diameter.

The TPM results actually depend on the dimensionless thermal inertia
$\Theta = \sqrt{\kappa\rho C (2\pi/P)}/(F_\odot /T_\circ)$ where $P$ is the rotation
period, $F_\odot$ is the flux from the Sun at the asteroid, and $T_\circ$ is the
equilibrium temperature of a flat surface facing the the Sun (Vokrouhlicky 1998).
The dimensionless thermal inertia $\Theta_1$ is defined for a distance from the Sun
of $r=1$ AU, and scaled with distance from the Sun as $\Theta = (r/[1 \mbox{AU}])^{0.75} \Theta_1$ 
to be consistent with the scaling used by Hung \etal\ (2022).

Small asteroids are often not spherical, and show optical
lightcurve variations that can be used to deduce the rotational period.
With accurate lightcurve data from multiple apparitions the rotation
pole and asteroid shape can be found.  Popescu \etal (2025) report
approximate pole and shape results on 2001 CC$_{21}$.

\begin{figure}[p]
\plotone{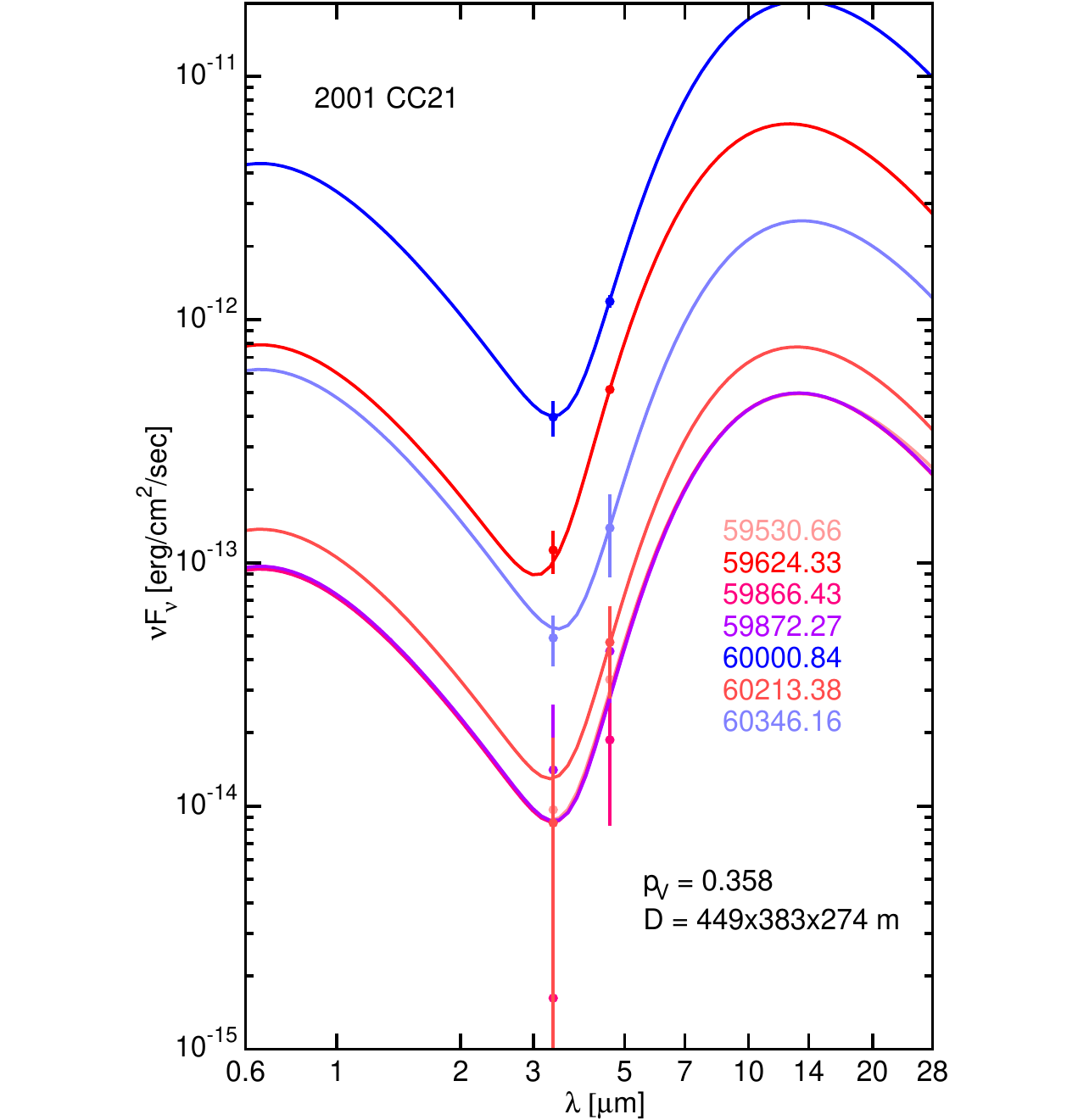}
\caption{The spectra of the ellipsoidal model with the highest posterior likelihood
compared to the NEOWISE data.\label{fig:ellipsoid-spect}}
\end{figure}

The infrared lightcurve data from NEOWISE are sparse and noisy, 
so the shape  derived from the IR data is not well determined.   
But an oblate ellipsoid will
be brighter when the rotation pole is pointing toward the Sun because it
will intercept more sunlight, and this will
lead to correlations between the rotation pole, shape, and thermal inertia.
Thus a non-spherical shape provides another possible explanation for 
the brightness differences between epochs that a spherical thermophysical model uses to
fix the rotation pole and thermal inertia.
This effect can be quantified by the ratio of the mean projected area
when viewed equator-on to the pole-on projected area which is 
$0.5(ac+bc)/ab = 0.5(c/b + c/a)$. This ratio is
0.72 using  Popescu \etal\ (2025), 
0.73 for the specific model in Figure \ref{fig:ellipsoid-spect},
and 0.73 using the medians in Table \ref{tab:ellipsoid}, so the
effects of shape could be substantial if the obliquity is $\sim 90^\circ$.

The three axes of the ellipsoidal model give two shape parameters: 
the axis ratios $b/a$ and $c/b$;
and the volume equivalent sphere diameter $2(abc)^{1/3}$.
The priors for the shape parameters are that $(b/a)^3$ and $(c/b)^4$ are
uniform in $[0..1]$.  
For $b/a$, which is easier to measure, the Thousand Asteroid Lightcurve
Survey (Masiero \etal\ 2009) showed that $p(b/a) \propto (b/a)^2$, so
$(b/a)^3$ has a uniform distribution.  This prior has a median $b/a$ of 0.8
For $c/b$ we assume that $(c/b)^4$ is uniform which means that
the $c/b$ ratio is less likely to be very different from 1.  The median
of this prior is $c/b = 0.85$.
While the median posterior shape in Table \ref{tab:ellipsoid} is not
significantly different from the median of the  prior, the posterior shape
distribution is somewhat narrower than the prior distribution.
The $c/b$ ratio is rather (anti-) correlated with the diameter as expected
from the discussion above.

\begin{figure}[p]
\plotone{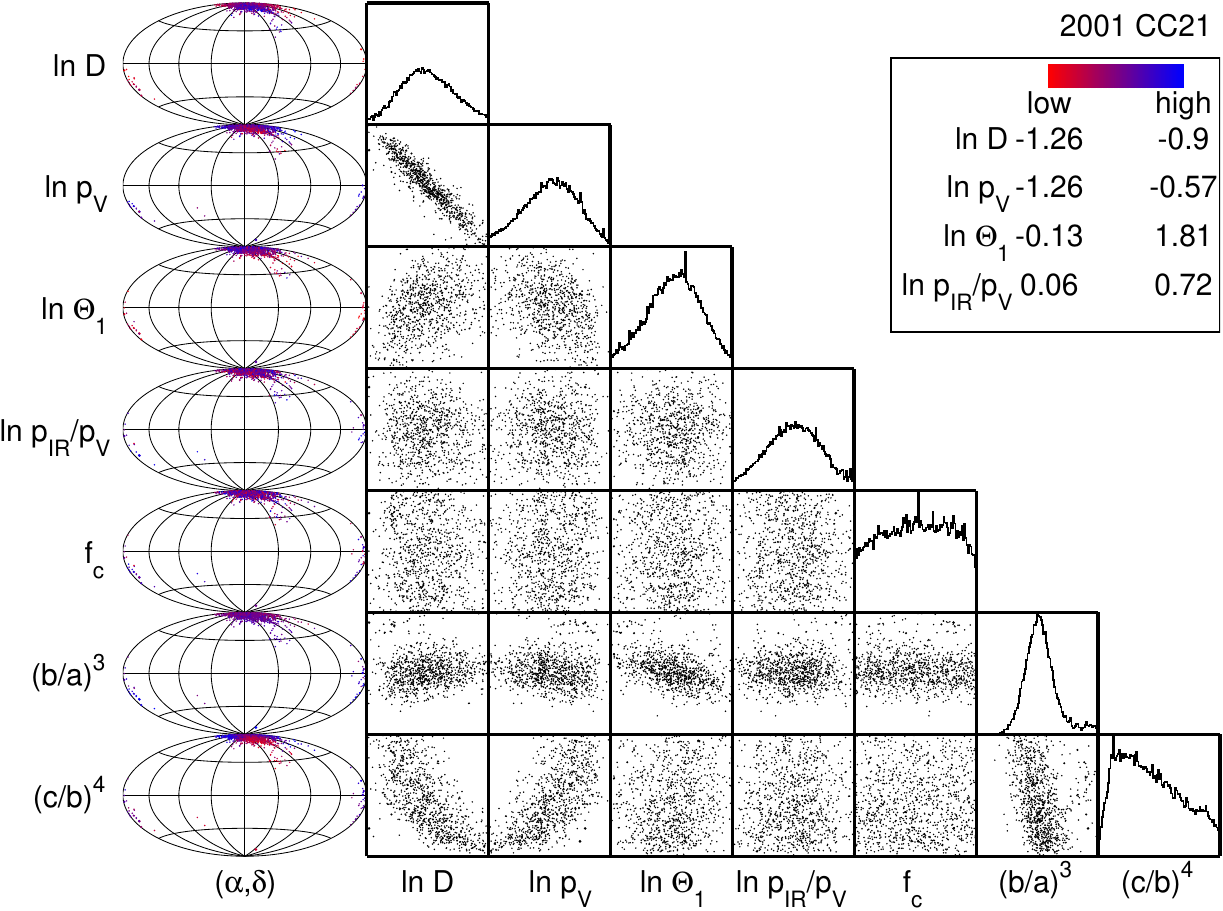}
\caption{Correlation between the parameters of the ellipsoidal TPM
model fit to the NEOWISE data.  
The maps and scatter diagrams show a random sample of
1000 models selected from the MCMC, while the histograms show
the entire chain.  The range for the scatter diagrams and histograms
cover a ``$\pm 2\sigma$'' range for each parameter that is actually
given by $(p_{16}+p_{84})/2 \pm (p_{84}-p_{16})$ where $p_{16}$ 
and $p_{84}$ are the $16^{1h}$ and $84^{th}$ percentile values
of the parameter $p$.  These ranges are shown in the upper right
corner.
However  the range is $[0..1]$ for $f_c$, $(b/a)^3$, or $(c/b)^4$.
The maps show the rotation poles of the models in ceelestial
coordinates.
The points on the maps are colored to show how a given parameter
correlates with the pole position, so a red dot on the map
shows a model with a low value of the parameter. \label{fig:ellipsoid-corner}}
\end{figure}

\begin{figure}[tb]
\plotone{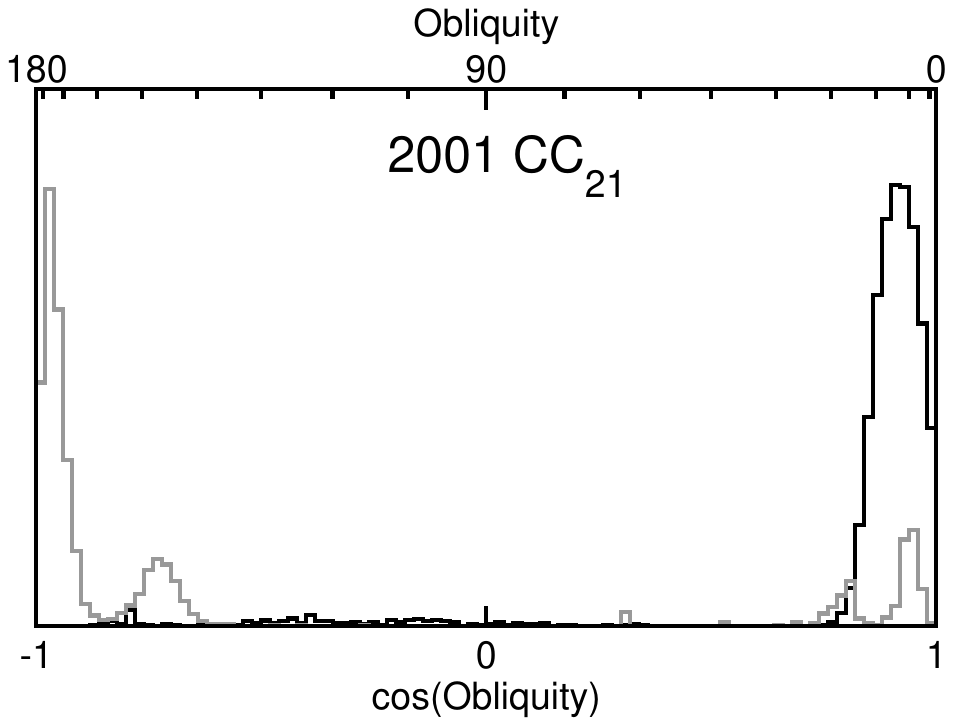}
\caption{Histogram of the cosine of the obliquity.
The prior would be uniform in these coordinates.
The black histogram shows the posterior distribution using the NEOWISE data alone..
The gray histogram shows the posterior when using both the NEOWISE and the
Spitzer data.
\label{fig:ellipsoid-obliquity}}
\end{figure}

The orientation of the rotation pole is given by two parameters: the
right ascension and declination.  The prior distribution for the rotation
pole is uniform in solid angle over $4\pi$.  The posterior distribution
can be seen in Figure \ref{fig:ellipsoid-corner}.  There is a peak 
near the North celestial pole at $(\alpha,\delta) = (259^\circ,87^\circ)$
and a few secondary clumps with retrograde rotation.
These  peaks can also be seen in Figure \ref{fig:ellipsoid-obliquity}
which shows a histogram of the cosine of the obliquity.  The prior would be
uniform in this variable.  The MCMC convergence to the correct weighting
of these  peaks is very slow, but the northern peak is clearly preferred.
The conditional distribution if the rotation is assumed to be prograde
gives the smaller obliquity range reported in the abstract. 

The surface roughness is not constrained by the available NEOWISE data
on 2001~CC$_{21}$, so the posterior
distribution of the crater fraction is uniform in $[0..1]$ as seen in
Figure \ref{fig:ellipsoid-corner}.  The surface roughness affects
the infrared phase function.

The ratio of the infrared albedo to the optical albedo is a parameter
of the model.  The central value of the prior on this ratio is taken from 
Table 1 in  Mainzer \etal (2011b) for NEOs while the scatter is taken from 
Masiero \etal (2014) for main belt asteroids.  The data do not strongly constrain
this value so the posterior is similar to the prior.

When parameters are not constrained by the data, the posterior distribution
samples over the prior, and any changes in the diameter correlated
with the unconstrained parameters are included in the final diameter
uncertainty.  A simple example illustrates this effect:  if the only datum is
the optical $H$ magnitude, then the median diameter estimate is given by
the median of the albedo prior, and the width of the albedo prior gives
the uncertainty in the diameter, which usually covers a factor of 2.5 or 3
between the 16$^{th}$ and 84$^{th}$ percentiles. 
 Any infrared data in thermally dominated bands
greatly reduces this diameter uncertainty.

There is very loose prior on the diameter, so the final size estimate
is based on the flux data which include both infrared fluxes and the
$H$ magnitude.  

The albedo $p_V$ is a parameter of the model, and it is determined by
balancing a prior based on hundreds of NEOs observed by WISE
(Wright \etal\ 2016) against the observed ratio of the optical to infrared flux
of the object.
The optical to infrared flux ratio of 2001~CC$_{21}$ is rather high
compared to other NEOs,  so if the assumed uncertainty in the
$H$ magnitude is increased giving the prior  more weight relative
to the data then the most likely albedo parameter will go down.  
When the albedo goes down, the temperature of the model goes
up like $(1-A)^{1/4}$, and the diameter needed to match the 
observed infrared flux goes down slightly as well.
But when the $H$ magnitude is well determined, as we assume here, 
the usual inverse correlation between  albedo and diameter
is found, which can be seen in the  $\ln(p_V)$ versus $\ln D$ scatter
diagram in Figure \ref{fig:ellipsoid-corner}.
The Fornasier \etal\ (2024) $H = 18.94 \pm 0.05$ is not used here,
but it would lead to a smaller albedo. .
But the $G_1 G_2$  phase function Fornasier \etal\ find gives a slightly  
higher phase integral ($q = 0.415$ versus $q = 0.384$ for $G=0.15$)
so the change in the Bond albedo $A = pq$ is quite small.
As a result the diameter would be almost unchanged.

The maximum posterior likelihood encountered during the MCMC is shown in
Figure \ref{fig:ellipsoid-spect}.  This model has a $\chi^2$ of 4.9
and a volume equivalent sphere diameter $2(abc)^{1/3}$ of 351 meters.
The axis ratios of this model are $b/a =  0.847$ and $c/b = 0.791$, 
similar to the prior median values.
The predicted $H$ magnitude for this model is 18.83 which is quite close
to the input.

The maximum posterior likelihood is not invariant under 
reparameterization of the model, so it is better to use the median and
$16^{th}$ to $84^{th}$ percentile ranges in Table \ref{tab:ellipsoid} 
for the TPM results and uncertainties.
For uncertainties given as a percentage, the reported values are
$\mbox{percent} = +100 \ln (84^{th}\mbox{percentile}/\mbox{median})$.
and $-100 \ln (\mbox{median}/16^{th}\mbox{percentile})$.

\section{Comparison to the Spitzer IRS results}

The diameter given by Fornasier \etal\ (2024) using the NEATM is 32\% larger
than the ellipsoidal TPM results reported here.  This is 3.35 times
the standard deviation of the difference.  Contributors to this discrepancy
include:
\begin{enumerate}
\item The NEATM requires color information to determine the beaming
parameter $\eta$ and the diameter, but the 4
different orders of the IRS spectrum had to be stitched together using
correction factors.  This practice is standard for the IRS to allow for
small  pointing  offsets that affect slit losses, but the corrections for
the 2001 CC$_{21}$ spectrum were larger due to rotational phase
effects.  These correction factors  should be carried as nuisance
parameters in the NEATM analysis, which 
would substantially increase the statistical error
on the diameter derived using the NEATM.
\item The shape of the infrared spectrum changes during the lightcurve,
so the stitched together spectrum is a patchwork quilt that need not
match the actual spectrum at any lightcurve phase.
\item Wright (2007) and Mommert \etal\  (2018) both compared NEATM fits 
to simulated data produced by a spherical TPM.
Wright (2007) found that the NEATM and a spherical TPM disagreed by
$\sim 8$\% RMS in diameter for Spitzer-visible NEOs observed between
50$^\circ$  and 60$^\circ$ phase angle when using 12 and 23 $\mu$m data.
Thus even with perfect color information and a perfect match between the
true shape and the modeled shape
 the error on the NEATM derived diameter
has to be much larger than the $\pm 3\%$ given by Fornasier \etal.
\end{enumerate}
If we take 10\% as a reasonable lower limit on the error for the Fornasier \etal\ diameter, the
discrepancy is $<2.4\sigma$.

Another reason to suspect the NEATM result is that this model does not conserve energy,
since the dayside temperature is reduced by a factor of $\eta^{1/4}$ but there is no 
nightside emission to compensate.  
Thus a fraction $(1-A)(1-1/\eta) \approx 0.4$ of the solar radiation
that hits the asteroid is lost.  
Models have been proposed to allow for this missing energy: for example
the Night Emission Simulated Thermal Model (NESTM, Wolters \& Green 2009),
but this has a minimal effect on the estimated diameter because
only a small fraction of the observable disk is on the nightside for a phase angle of $53^\circ$.

The relatively high  beaming parameter $\eta = 1.847 \pm 0.034$ found by Fornasier \etal\ (2024)
translates into a lower limit on the thermal inertia, $\Gamma > 370\;\mbox{MKS}$ 
using the correlation in Harris \& Drube (2016),
which agrees well with our prior on $\Gamma$ based on Hung \etal\ (2022).
This beaming parameter is well within the 16$^{th}$ to 84$^{th}$ percentile range of 
$\eta$'s found by Mainzer \etal\ (2014a) for small NEOs using fits to 4-band WISE data.

We can estimate the accuracy and precision of the NEATM by applying our TPM to the Spitzer
observation.   Figure A.2 of Fornasier \etal\ (2024) shows 8 spectra obtained with the
Spitzer IRS
Short wavelength Low resolution section, the Long wavelength Low resolution
section, in grating orders 1 and 2, and in two dithers a and b.  
We chose wavelengths near the center of each order that were already
calculated by the TPM, and thus measured fluxes from Figure A.2 at 6.295, 10.353,
17.894 and 28 $\mu$m.  The TPM computes both the mean flux and the $\sin$ and $\cos$
components of the lightcurve approximated as a sinusoid with two cycles per rotation.
Both the amplitude and the phase of the sinusoid depend on the wavelength.

The SL1 and SL2 dithers were only 
a few minutes apart
in time, so we just average the a and b dithers.  
The dithers were 25 minutes apart for LL2 and 50 minutes apart
for LL1.  The flux went down between LL2a and LL2b and up
between LL1a and LL2b.  
We compare these measured fluxes which have a known separation in
time with the sinusoidal approximations to the lightcurve computed
by the TPM.
Since we do not use the absolute rotational phase in our model,
we introduce a new parameter, the rotational phase
$\phi$ for the first observation, and evaluate $\chi^2$ for 60 steps in
$\phi$.  The best $\chi^2$ is used in the likelihood for the Monte Carlo
Markov Chain calculation.  In computing $\chi^2$ we assume a 10\% flux
uncertainty for each of the 6 values.

With 6 data points and 11 free parameters that are all constrained by priors,
the median posterior and $16^{th}$ to $84^{th}$ percentile range for the
diameter is $476^{+9.5\%}_{-8.5\%}$ meters.  This essentially matches the
NEATM estimate  with errors that are close to $\pm10\%$ as we expected.
The TPM diameter estimate computed only from the Spitzer data is shown 
as the magenta errorbar on Figure \ref{fig:dhist}.

The thermal inertia posterior distribution for this fit only to the Spitzer data is
$\Gamma = 748^{+39\%}_{-38\%}$, while the Hung \etal\ prior for the
median diameter is $\Gamma = 419 \pm 46\%$, so the Spitzer data alone
favors a thermal inertia that is slightly (1.26$\sigma$) higher than the prior.

The posterior distribution of the shape parameters give
$b/a = 0.565^{+0.157}_{-0.192}$ which is more
elongated than the prior, since $(b/a)^3 = 0.18$.  The shape is also
slightly flatter than the prior with $(c/b) = 0.755^{+0.201}_{-0.303}$.

The posterior distribution of the obliquity is quite broad: a median of $53^\circ$
with a 16$^{th}$-84$^{th}$ percentile range of 16$^\circ$ to 152$^\circ$.  There
is a slight preference for prograde rotation because that means Spitzer was
looking at the morning side of the object which is naturally cooler without
requiring an even higher thermal inertia.	

\section{Joint Spitzer-NEOWISE modeling}

\begin{figure}[p]
\plotone{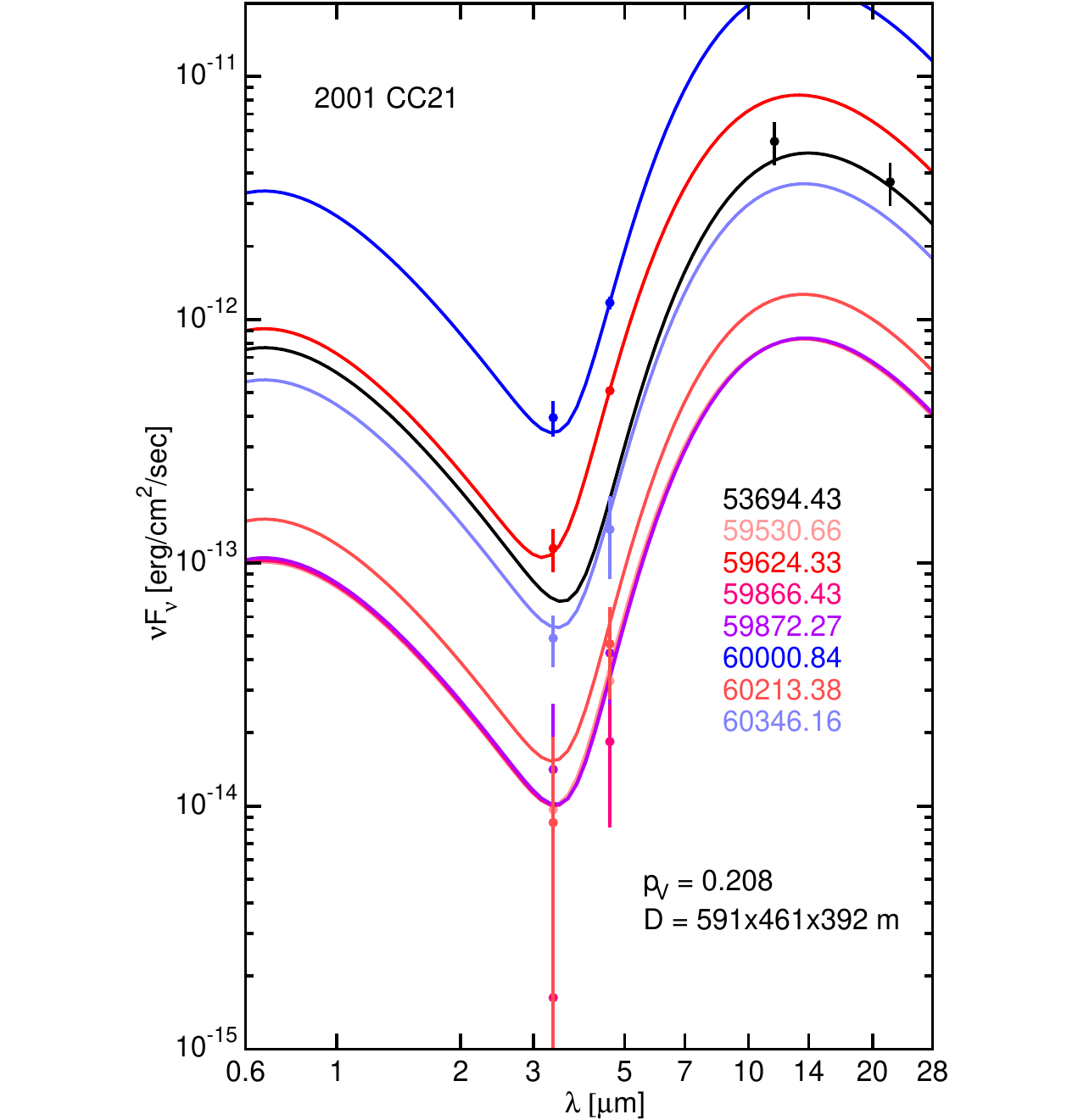}
\caption{Spectral energy distributions for the Spitzer observation and the
many epochs of NEOWISE observations predicted by the maximum posterior model
found during the MCMC.  The thermal inertia for this model
is $\Gamma_1 = 2111$ MKS units.  The rotation pole is at $(\alpha, \delta) =
(281^\circ,  45^\circ)$.  
The diameter of a sphere with the same volume is $D = 2(abc)^{1/3} = 474$ meters.
\label{fig:2001CC21-SED-w-Spitzer}}
\end{figure}

The Spitzer data presented by Fornasier \etal\ provide measurements around the peak
of the spectral energy distribution of 2001 CC$_{21}$ which allows for a more reliable 
determination of the bolometric flux from the asteroid.  The modeling based on the NEOWISE
data alone has to extrapolate by a factor of ~20 to determine the peak of $\nu F_{\nu}$,
which introduces extra uncertainty into the NEOWISE result.
A blackbody falls by a factor of 20 below the peak at $h\nu/kT \approx 11$, so the 
surface brightness is very dependent on the temperature.  For an asteroid model
with a range of temperatures, the surface brightness in the Wien tail is highly
dependent on the hottest spots in the model, which are the crater bottoms at the
subsolar point in the rotating cratered thermophysical model.   But these are not
not visible at the phase angles observed here, so the crater fraction parameter
has little effect on the model fluxes.
However, the surface brightness
at the peak of $\nu F_\nu$ is always close to the incident solar flux and the
diameter determination is much more certain.  We take advantage of the
Spitzer data near the peak of the spectral energy distribution by
adding the $\chi^2$ from the Spitzer data discussed above to the many 
epochs of NEOWISE data, and repeated the Monte Carlo Markov Chain
calculation.
The best fit to the data found
while constructing the chain is shown in
Figure \ref{fig:2001CC21-SED-w-Spitzer}.  
We have plotted the Spitzer data as pseudo-WISE data
by averaging the two dithers in the LL1 and LL2 orders,
and then interpolating to the WISE W3 and W4 wavelengths.  Due to the width of the W3
filter, a color correction taken from Wright \etal\ (2010) was applied.
However, the actual fit was to the 6 mid-order fluxes as described in the text above.
While this is formally an acceptable fit with $\chi^2 = 13.2$ when fitting 11 parameters to 22 data points,
the thermal inertia for this model is $\Gamma_1 = 2111$ MKS units
which is much higher than the
thermal inertia prior based on Hung \etal\ (2022),  As a result large diameter models like
this were disfavored in a fit to the NEOWISE data without Spitzer.

Note that the thermal inertia prior enters the likelihood calculation as a (previously
known) datapoint.  Thus large deviations like this ($> 2\sigma$) are downweighted.
Since there are 7 NEOWISE 4.6 $\mu$m fluxes and 6 Spitzer flux values, the
data dominate the prior is this case.  With the large fitted value for the thermal inertia, the
model approaches the fast rotating model (FRM) limit.  In the FRM limit, prograde \vs\
retrograde rotations cannot be distinguished.

\begin{figure}[p]
\plotone{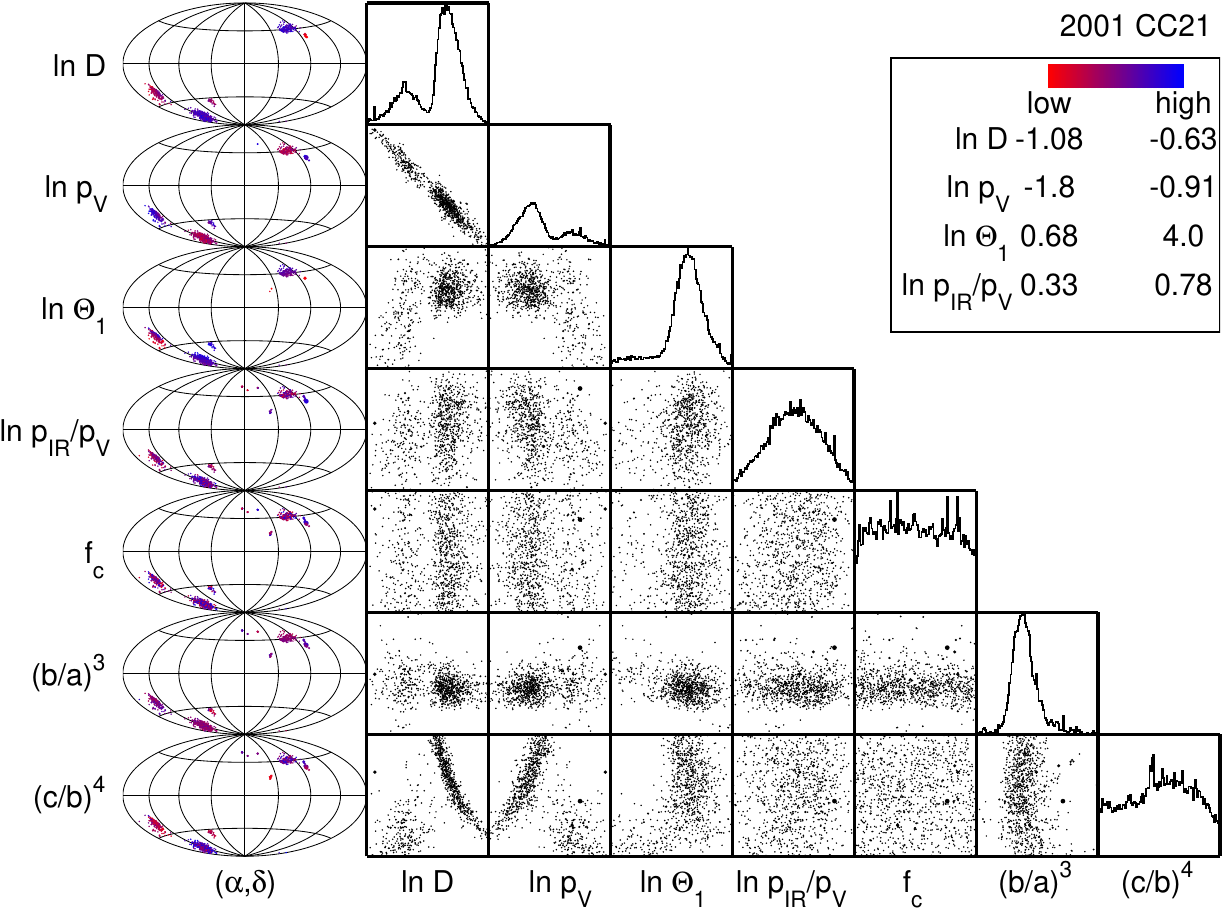}
\caption{Correlation between the parameters of the ellipsoidal TPM
model fit to both the NEOWISE and Spitzer data.  
See Figure \ref{fig:ellipsoid-corner} for an explanation of the plot.
The obliquity distribution has a median of 160$^\circ$
with a 16$^{th}$-84$^{th}$ percentile range of $38^\circ$ to 168$^\circ$
The thermal inertia is much higher than the prior, with
$\Gamma = 2924^{+39\%}_{-105\%}$ MKS.
\label{fig:ellipsoid-corner-w-Spitzer}
}
\end{figure}

The parameter correlations are shown in Figure \ref{fig:ellipsoid-corner-w-Spitzer}.
There are strong correlations between parameters, most notably between the
diameter, the oblateness $c/b$, and the rotation pole.  
The contrast between Figure \ref{fig:ellipsoid-corner} and Figure
\ref{fig:ellipsoid-corner-w-Spitzer} is dramatic.
The NEOWISE data
were taken from a wide range of angles but have low SNR.
The Spitzer data have very limited angular coverage but cover the
peak of the spectral energy distribution.
These limitations of the data allow the model to trade off physical
parameters against geometric parameters.
This contributes to the diameter bimodality seen in Figure  \ref{fig:dhist}.
One such tradeoff changes the prograde rotation preferred by the NEOWISE
data to retrograde rotation which helps to explain the high flux seen by Spitzer,
since for retrograde rotation Spitzer observed the afternoon side of
the object.  The posterior distribution of the cosine of the obliquity
shown in gray in Figure \ref{fig:ellipsoid-obliquity} is multimodal.
This is a projection of the two dimensional rotation pole maps seen
in Figure \ref{fig:ellipsoid-corner-w-Spitzer} where the multiple islands
of acceptable fits 
cause slow convergence of the Monte Carlo Markov chain.

Using the Popescu \etal\ (2025) solution for a shape model and rotation pole from
optical lightcurve data would fix these geometric degrees of freedom
and allow a more precise size estimate using the infrared fluxes.
M\"uller \etal\ (2014) show how this technique can reach 2\% accuracy
on diameters.

\clearpage

\section{Discussion}

\begin{figure}[p]
\plotone{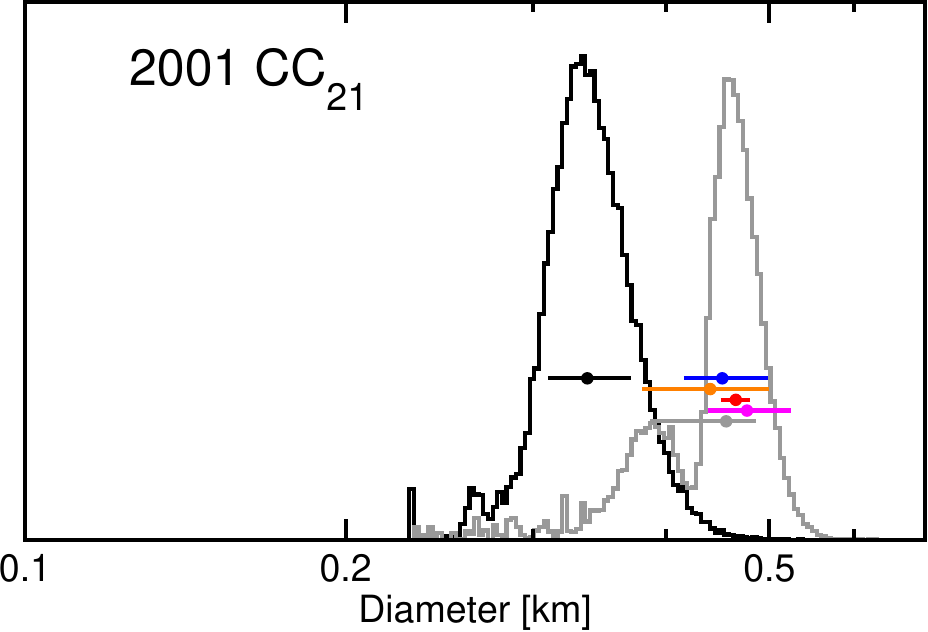}
\caption{The posterior distribution of the equivalent volume sphere
diameter from the ellipsoidal model applied to the NEOWISE data (black), 
with the 16$^{th}$ to 84$^{th}$ percentile
range indicated by a black horizontal errorbar; the
value derived from the polarimetric 
albedo (Geem \etal\ 2023)  indicated by the blue errorbar; 
the shape model of Popescu \etal\ (2025) scaled using the
Arimatsu \etal\ (2024) occulation data as an orange errorbar; and
the Fornasier \etal\ (2024) NEATM result indicated by the red errorbar.
The magenta errorbar shows the result of the TPM applied to the
Fornasier \etal\ data.
The grey histogram and errorbar shows the TPM applied to the
Fornasier \etal\ data combined with the NEOWISE data.
The Hirabayashi \etal\ (2021) value of 700 m would be at the right edge
of the graph.
\label{fig:dhist}}
\end{figure}

We have presented an analysis of the thermal IR data from NEOWISE
on 2001 CC$_{21}$ taken from 2021 to  2024.  A thermophysical analysis
of these data suggest a diameter much less than the 700 meters in
Hirabayashi \etal\ (2021), and therefore a high albedo that is
somewhat higher than the albedo derived from polarimetry by Geem
\etal\ (2023).  
We also have used the TPM to analyze the Spitzer data presented by Fornasier \etal\ (2024).
When using the TPM
we find that the NEOWISE data, the Spitzer data, and the Hung \etal\ (2022)
prior on the thermal inertia are mutually inconsistent at the 2-3$\sigma$ level,
like 3 lines of position in celestial navigation that cross in a triangle instead of a
point.
Figure \ref{fig:dhist} shows the range of diameters that are consistent
with the TPM values presented here and the results from Geem \etal\ (2023)
and Fornasier \etal\ (2024).
The  ellipsoidal TPM model  fit only to the NEOWISE data gives a diameter that is 34.5\%
lower than the TPM model diameter using only the Spitzer data.  If we consider three distinct
measurements of the diameter (Spitzer, NEOWISE \& polarimetry) with each having a 
Gaussian distribution with a 10\% 
standard deviation, the probability
of a deviation this large is 2.7\%, corresponding to a 1.93$\sigma$ effect.
The actual black histogram of the NEOWISE-only diameters in Figure \ref{fig:dhist} has an excess kurtosis of 1.18 
which indicates heavier than Gaussian tails, so 2.7\%  is a lower limit on the probability
that the discrepancy is just a fluke. 

Thus this deviation could be a statistical fluctuation, but the preponderance of the
data indicates a real anomaly  such as a
$\approx 470$ meter object with an unexpectedly high thermal inertia.  
Such an anomaly could be a
peculiarity of 2001 CC$_{21}$ or an indication that the priors we have used should be
adjusted.
We note that the Hung \etal\ (2022) dataset included hundreds of objects
larger than a few km, but very few objects smaller than 1 km. 
These include spacecraft observed NEOs:
$\Gamma = 750$ for Itokawa (Fujiwara \etal\ 2006), 200-500 for Ryugu, taken as 316 (Sugita \etal\ 2019), and
$310$ for Bennu (Rozitis \etal\ 2020).  These well-determined $\Gamma$ values differ from the centerline
of our prior by $+44\%$, $-6\%$, and $-29\%$ respectively, roughly consistent with the
prior, but $n=3$ is a very small sample size and we cannot rule out a fat tail.

Another possible anomaly is a substantial dip ( $\approx 50\%$) in the emissivity in the 4-5 \um\ band.
The Fornasier \etal\ (2024) Spitzer IRS data were taken to search for emissivity features
and nothing close to this large was seen, but the Spitzer IRS does not cover the 4-5 \um\ band.
However, a reduced emissivity is seen at millimeter wavelengths for Mars by WMAP
(Weiland \etal\ 2011), and for asteroids by the Atacama Cosmology Telescope
(Orlowski-Scherer \etal 2024).  But the TPM model assumes $\epsilon = 0.95$
at all wavelengths so there are no emissivity parameters in the model.

While the TPM has a large number of parameters, some of which are not constrained
by the NEOWISE data, this is in fact  the usual case when estimating asteroid diameters.
For example, to convert a measured optical flux into a diameter requires an albedo and
a phase function shape which introduces two parameters that must be fixed by priors..  
The usual assumed phase curve takes $G = 0.150$ based on prior knowledge.  The albedo is quite
uncertain with a bimodal prior distribution 
(Wright \etal\ 2016) having a median at 0.151 but with the 16$^{th}$
and 84$^{th}$ percentiles at 0.038 and 0.295, a factor of 7.7 range.

The imagery from the Hayabusa2$\sharp$ flyby in July 2026 will provide the ground truth
to assess the accuracy of the diameter estimates presented in this paper, in Geem \etal\ (2023),
in Arimatsu \etal\ (2024),
and in Fornasier \etal\ (2024).  This will provide a valuable calibration of size estimation
techniques on objects smaller than a kilometer.  In addition, resolved thermal IR imaging could
indicate whether the thermal inertia is really as high as our joint fit of NEOWISE plus
Spitzer IRS suggests. 

\begin{acknowledgements}

\noindent ACKNOWLEDGEMENTS

This work was supported in part by a NASA contract to UCLA; 80MSFC24CA008.
This publication makes use of data products from the Wide-field Infrared Survey Explorer, 
which is a joint project of the University of California, Los Angeles, and the Jet Propulsion 
Laboratory/California Institute of Technology, funded by the National Aeronautics and 
Space Administration. This publication also makes use of data products from NEOWISE, 
which is a project of the Jet Propulsion Laboratory/California Institute of Technology
and UCLA,
funded by the Planetary Science Division of the National Aeronautics and Space Administration.

This research has made use of the NASA/IPAC Infrared Science Archive, which is 
funded by the National Aeronautics and Space Administration and operated by the 
California Institute of Technology.

This research has made use of data and/or services provided by the International 
Astronomical Union's Minor Planet Center.

Data set usage:

NEOWISE-R Single-Exposure Images (NEOWISE-R Team 2020).

\end{acknowledgements}

\facilities{IRSA, NEOWISE}

\clearpage

\section{References}

\parskip=10pt plus 1pt

\noindent  Arimatsu, K., Yoshida, F., Hayamizu, T., et al.\ 2024, \pasj, 76, 940. doi:10.1093/pasj/psae060

\noindent Bowell, E., Hapke, B., Domingue, D., Lumme, K., Peltoniemi, J. \& Harris, A.,
1989, in Asteroids II  (ed. Binzel, R., Gehrels, T. \& Matthews, M.),  524-556

\noindent Delbo M, Mueller M, Emery JP, Rozitis B, Capria MT. 2015. Asteroid Thermophysical Modeling. in Asteroids IV
107-128
 
 \noindent Emery, J.~P., Fern{\'a}ndez, Y.~R., Kelley, M.~S.~P., et al.\ 2014, \icarus, 
 234, 17. doi:10.1016/j.icarus.2014.02.005
 
 \noindent Fornasier, S., Dotto, E., Panuzzo, P., et al.\ 2024, \aap, 688, L7. doi:10.1051/0004-6361/202450447
 
 \noindent Fujiwara, A., Kawaguchi, J., Yeomans, D.~K., et al.\ 2006, Science, 
 312, 5778, 1330. doi:10.1126/science.1125841
 
 \noindent Geem, J., Ishiguro, M., Granvik, M., \etal\ 2023,  MNRAS, 525, L17. doi:10.1093/mnrasl/slad073
 
 \noindent Harris, Alan W. 1998, Icarus, 131, 291.

\noindent Hasegawa, S., M{\"u}ller, T.~G., Kawakami, K., et al.\ 2008, \pasj, 60, S399. doi:10.1093/pasj/60.sp2.S399

\noindent Hirabayashi, M., Mimasu, Y., Sakatani, N., et al.\ 2021, 
Advances in Space Research, 68, 1533. doi:10.1016/j.asr.2021.03.030

\noindent Hung, D., Hanu{\v{s}}, J., Masiero, J.~R., et al.\ 2022, 
Planetary Science Journal, 3, 56. doi:10.3847/PSJ/ac4d1f

\noindent  Mainzer, A., Grav, T., Masiero, J., \etal, 2011a, ApJ, 741, 90. doi:10.1088/0004-637X/741/2/90

\noindent Mainzer, A., Grav, T., Bauer, J., et al.\ 2011b, \apj, 743, 156. doi:10.1088/0004-637X/743/2/156

\noindent Mainzer, A., Bauer, J., Grav, T., et al.\ 2014a, \apj, 784, 110. doi:10.1088/0004-637X/784/2/110

\noindent Mainzer, A., Bauer, J., Cutri, R.~M., et al.\ 2014b, \apj, 
792, 1, 30. doi:10.1088/0004-637X/792/1/30

\noindent Masiero, J., Jedicke, R., {\v{D}}urech, J., et al.\ 2009, Icarus, 204, 145. doi:10.1016/j.icarus.2009.06.012

\noindent Masiero, J.~R., Grav, T., Mainzer, A.~K., et al.\ 2014, ApJ, 791, 121. doi:10.1088/0004-637X/791/2/121

\noindent Mommert, M., Jedicke, R., \& Trilling, D.~E.\ 2018, \aj, 155, 74. doi:10.3847/1538-3881/aaa23b

\noindent M{\"u}ller, T.~G., Hasegawa, S., \& Usui, F.\ 2014, \pasj, 
66, 3, 52. doi:10.1093/pasj/psu034

\noindent NEOWISE-R Team, 2020, NEOWISE-R L1b Images, IPAC, doi:10.26131/IRSA147

\noindent Orlowski-Scherer, J., Venterea, R.~C., Battaglia, N., et al.\ 2024, \apj, 
964, 2, 138. doi:10.3847/1538-4357/ad21fe

\noindent Popescu, M.~M., Tatsumi, E., Licandro, J., et al.\ 2025, \psj, 6, 42. doi:10.3847/PSJ/ada560

\noindent Rozitis, B., Ryan, A.~J., Emery, J.~P., et al.\ 2020, Science Advances, 
6, 41, eabc3699. doi:10.1126/sciadv.abc3699

\noindent Sugita, S., Honda, R., Morota, T., et al.\ 2019, Science, 
364, 6437, eaaw0422. doi:10.1126/science.aaw0422

\noindent Vokrouhlicky, D. 1998, A\&A, 335, 1093.


\noindent Watanabe, S., Hirabayashi, M., Hirata, N., et al.\ 2019, Science, 
364, 6437, 268. doi:10.1126/science.aav8032

\noindent Weiland, J.~L., Odegard, N., Hill, R.~S., et al.\ 2011, \apjs, 
192, 2, 19. doi:10.1088/0067-0049/192/2/19

\noindent Wolters, S.~D. \& Green, S.~F.\ 2008, Asteroids, Comets, Meteors 2008, 1405, 8120

\noindent Wright, E. L. 2007, arxiv.org/astro-ph/0703085v2
 
\noindent  Wright, E.~L., Mainzer, A., Masiero, J., Grav, T. \& Bauer, J. \ 2016, \aj, 152, 4, 79. doi:10.3847/0004-6256/152/4/79

\noindent  Wright, E., Masiero, J., \& Mainzer, A.\ 2023, \dps

\clearpage

The full Table 1:
\begin{verbatim}
    MJD.      Flux[DN]  MAD[DN] Frame ID  Notes
59529.44243     32.6      2.9  33634r135
59529.44256      0.9      2.6  33634r136
59529.57321     31.1      2.9  33638r135
59529.57333      3.0      3.0  33638r136
59529.70398     59.7      3.1  33642r136
59529.83463      4.4      2.7  33646r136
59529.96528     14.8      3.3  33650r135
59530.09606      3.7      3.1  33654r136
59530.22670     47.5      3.0  33658r136
59530.35735     46.4      2.9  33662r136
59530.42268     88.8      2.5  33664r111
59530.42281     66.4      3.3  33664r112
59530.48762     87.4      4.7  33666r108
59530.55294    -32.6      3.1  33668r055
59530.61878     31.6      3.1  33670r136
59530.68410      6.7      3.1  33672r111
59530.74943     43.9      3.0  33674r136
59530.81475     27.1      3.2  33676r111
59530.88008    -14.7      2.7  33678r135
59530.94553     21.8      2.6  33680r111
59531.01085     17.0      2.8  33682r136
59531.07618     -4.8      4.5  33684r111
59531.20632     36.4      3.7  33688r067
59531.33709     20.5      3.1  33692r115
59531.46774     49.5      2.6  33696r114
59531.59839     55.5      3.4  33700r112
59531.72904    -14.9      2.3  33704r111
59531.72917     77.5      2.7  33704r112
59531.85981     25.6      3.1  33708r112
59531.99046     19.9      2.8  33712r112

59623.71409    361.3      2.4  36520r081
59623.84462    280.4      2.7  36524r080
59623.84474    316.4      2.7  36524r081
59623.97526    261.4      3.5  36528r080
59624.10591    379.2      2.8  36532r080
59624.10604    415.1      2.7  36532r081
59624.17124    199.3      2.7  36534r105
59624.23656    388.8      2.7  36536r080
59624.30189    343.6      3.4  36538r105
59624.36721    221.2      2.0  36540r080
59624.43254    499.5      2.9  36542r105
59624.49786    215.9      3.8  36544r090
59624.56319    269.6      2.4  36546r047
59624.69383    256.2      2.8  36550r105
59624.82448    381.1      3.6  36554r106
59624.95500    402.0      2.4  36558r105
59624.95513    428.6      2.9  36558r106

59865.34260    188.2      4.7  43926r143
59865.47299      7.1      2.5  43930r143
59865.60339     21.5      2.7  43934r143
59865.73378     -8.2      2.8  43938r143
59865.86405     50.5      3.7  43942r142
59865.99444    -31.8      3.8  43946r142
59866.12484     59.0      2.9  43950r090
59866.19003     11.6      2.5  43952r058
59866.25523     28.3      3.0  43954r143
59866.32043    -22.5      2.1  43956r118
59866.38563     -9.0      3.0  43958r143
59866.45082      9.8      3.4  43960r118
59866.51602     -4.4      3.6  43962r143
59866.58122     -1.1      3.2  43964r118
59866.64629     20.4      4.0  43966r142
59866.64641    -11.0      3.4  43966r143
59866.71148    -10.5      3.6  43968r117
59866.84188     66.4      4.9  43972r117
59866.97227     17.1      3.8  43976r118
59867.10267    -33.0      4.3  43980r118
59867.23306     13.6      4.6  43984r118
59867.36345     49.0      4.8  43988r118
59867.49372     45.5      4.2  43992r091

59871.20982     70.6      3.5  44106r141
59871.34022     -5.6      2.9  44110r141
59871.47061     53.7      2.8  44114r141
59871.60088     42.8      3.3  44118r141
59871.73127    -30.4      3.6  44122r141
59871.86167     11.7      3.5  44126r141
59871.99206     32.5      7.4  44130r141
59872.05726    -13.5      3.7  44132r117
59872.12233    613.7     23.3  44134r140 D
59872.12245   1180.6     16.3  44134r142 D
59872.18714     24.6      3.0  44136r054
59872.25272     37.2      3.0  44138r141
59872.31792    105.8      3.0  44140r116
59872.38311    -19.4      2.7  44142r141
59872.44780    -28.6      3.4  44144r100
59872.51351     45.7      2.9  44146r141
59872.57870     35.1      3.2  44148r116
59872.70910     53.9      3.0  44152r117
59872.83937     15.9      2.5  44156r116
59872.83949     18.3      3.1  44156r117
59872.96976     -6.4      3.2  44160r116
59873.10015     18.1      2.9  44164r116
59873.23055     32.4      2.7  44168r116

60000.07929    874.3      2.9  48063r199
60000.07942    913.0      2.3  48063r200
60000.20943    635.3      2.8  48067r199
60000.27450    883.8      3.9  48069r175
60000.33957    613.5      2.7  48071r200
60000.40464    746.5      2.9  48073r176
60000.53478    465.4      3.1  48077r176

60212.38129    108.7      2.6  54606r147
60212.51079     19.2      3.6  54610r091
60212.64029    -19.9      3.0  54614r147
60212.76980     -1.7      5.1  54618r147
60212.89930     41.8      4.0  54622r147
60213.02880    140.1      3.6  54626r147
60213.15831    106.2      3.2  54630r068
60213.22299     20.1      2.4  54632r122
60213.28781    -30.9      3.6  54634r147
60213.35250     10.8      3.5  54636r122
60213.41731     15.4      4.2  54638r148
60213.48200     19.1      3.7  54640r122
60213.54669     40.1      4.9  54642r147
60213.54681      3.0      4.2  54642r148
60213.61150    190.2      8.2  54644r122
60213.74100     26.7      3.1  54648r122
60213.87051     34.6      3.4  54652r122
60214.00001     50.6      2.6  54656r123
60214.12938     12.6      2.3  54660r122
60214.12951     -4.7      3.3  54660r123
60214.25889     54.4      3.2  54664r122

60345.62653    120.1      2.4  58731r154
60345.75552     85.9      2.3  58735r155
60345.88439     52.6      2.7  58739r154
60346.01326    180.1      2.7  58743r154
60346.07769     50.2      2.9  58745r130
60346.14212    102.8      2.6  58747r154
60346.14225     94.1      2.7  58747r155
60346.20668    189.7      2.8  58749r130
60346.27112     33.6      2.8  58751r155
60346.33555     52.0      3.1  58753r130
60346.46442    112.3      3.0  58757r130
60346.59328     56.9      3.5  58761r130
60346.59341     59.4      3.6  58761r131
\end{verbatim} 
\end{document}